# Few-shot Class-incremental Audio Classification Using Stochastic Classifier

*Yanxiong Li, Wenchang Cao, Jialong Li, Wei Xie, Qianhua He*

School of Electronic and Information Engineering, South China University of Technology, Guangzhou, China
eeyxli@scut.edu.cn, eewenchangcao@mail.scut.edu.cn

## Abstract

It is generally assumed that number of classes is fixed in current audio classification methods, and the model can recognize pre-given classes only. When new classes emerge, the model needs to be retrained with adequate samples of all classes. If new classes continually emerge, these methods will not work well and even infeasible. In this study, we propose a method for few-shot class-incremental audio classification, which continually recognizes new classes and remember old ones. The proposed model consists of an embedding extractor and a stochastic classifier. The former is trained in base session and frozen in incremental sessions, while the latter is incrementally expanded in all sessions. Two datasets (NS-100 and LS-100) are built by choosing samples from audio corpora of NSynth and LibriSpeech, respectively. Results show that our method exceeds four baseline ones in average accuracy and performance dropping rate. Code is at https://github.com/vinceasvp/meta-sc.
**Index Terms**: few-shot learning, incremental learning, audio classification, stochastic classifier, prototype vector

## 1. Introduction

Audio classification aims to recognize different types of sounds in environment, which is a hot topic with wide applications, such as multimedia analysis [1], audio captioning [2], traffic surveillance [3], bio-acoustic monitoring [4], automatic assisted driving [5], and smart home [6].

Some works were done on audio classification in recent years [7]-[11]. In these works, it was generally assumed that the vocabulary of audio classes was fixed and pre-known. Hence, the trained model can recognize the pre-given classes only. To recognize new classes, the model needs to be retrained with abundant labeled samples of all classes. Retraining the model is quite time-consuming and laborious. If the samples of old classes are unavailable due to data privacy and memory shortage, finetuning the model with samples of new classes will make the model quickly forget old ones (catastrophic forgetting problem [12]). However, in many applications, the vocabulary of audio classes dynamically expands. For example, users of intelligent audio devices (e.g., smart speaker) often want to add new classes, such as abnormal sounds, or audio wake-up words.

To reduce the requirement for the amount of training samples, the methods of few-shot sound classification (recognition or detection) were proposed [13]-[18]. In these methods, the model can recognize new classes from a few training samples [19], [20], but they cannot remember old classes. To continually recognize new classes and remember old ones, the incremental learning [21], [22] based methods for audio classification (or detection) were proposed [23]-[27]. Although these methods can recognize new and old classes, they still have shortcomings. For instance, abundant samples of new classes are required to update the model. This requirement is difficult to meet in practice, since the labeled samples of some new classes are few.

Recently, a learning paradigm, Few-Shot Class-Incremental Learning (FSCIL) was proposed [28], [29], which aims to continually expand the model for recognizing new classes with few training samples and meanwhile remembering old ones. The FSCIL faces two main problems: overfitting to new classes because of very few samples of new classes, and catastrophic forgetting of old classes due to absence of samples of old classes in incremental sessions. To address these two problems, a decoupled learning strategy was proposed, in which the feature extractor and the classifier (two major components of the model) were trained independently [28]-[34]. Wang et al. applied the Dynamical Few-Shot Learning (DFSL) [35] to audio classification, which is the latest work related to the Few-shot Class-incremental Audio Classification (FCAC). Although these works above have merits, they still need to be improved for tackling the two problems faced by the FSCIL and FCAC. For example, their classifier consists of prototypes (a prototype represents a class) and each prototype is a deterministic vector. Each deterministic prototype vector might not represent the new class effectively, since it is learnt from few training samples of one new class. Hence, the classifier consisting of deterministic prototype vectors (called deterministic classifier here) might not obtain discriminative decision boundaries over all classes, and would not perform well for audio classification.

Inspired by the success of Stochastic Classifier (SC) in computer vision [36], we propose a FCAC method using a SC here. The SC is trained for generating prototypes and each prototype is sampled from a distribution which is determined by mean vectors and variance vectors. Each prototype is a mean vector with a margin (variance vector). Hence, many prototypes around the mean vector can be obtained and at least one of these prototypes is expected to represent the new class well. As is done in prior works, the embedding extractor is frozen in incremental sessions after training in base session, while the SC is dynamically expanded in incremental sessions for continually recognizing new classes. Two datasets, named NS-100 and LS-100 are constructed by choosing samples from two public audio corpora of the NSynth and LibriSpeech, respectively. To reproduce our experiments, the construction details (such as metadata, explanations) of these two datasets are introduced at https://github.com/vinceasvp/meta-sc. Results indicate that our method outperforms four state-of-the-art methods in terms of Average Accuracy (AA) and Performance Dropping rate (PD).

In conclusion, the main contributions of this study are as follows. First, we design a SC which is expanded in incremental sessions and is adopted to generate prototypes for continually recognizing new classes. Second, we propose a FCAC method whose effectiveness is evaluated on two audio datasets.

## 2. Method

In this section, we describe our method in detail, including problem definition, the framework of our method, and SC.

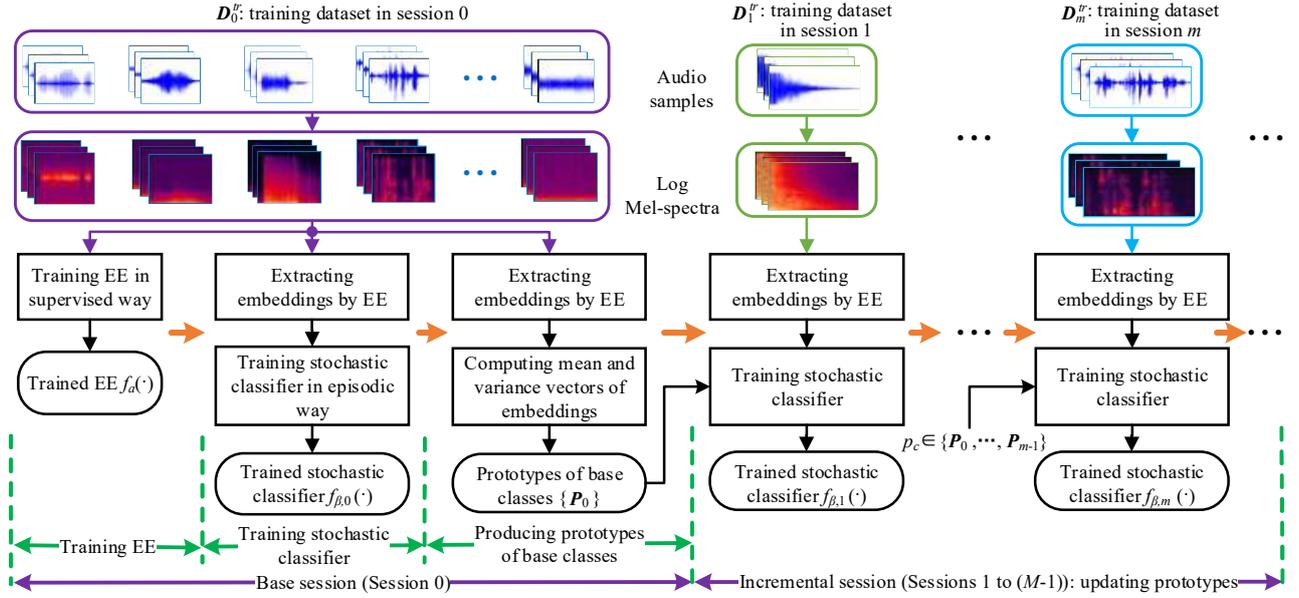

Figure 1: *The framework of our method consists of two kinds of sessions: base and incremental sessions. EE: Embedding Extractor.*

## 2.1. Problem definition

The FCAC includes $M$ sessions, namely one base session (session 0) and $M$-1 incremental sessions (sessions 1 to $M$-1). The training and testing datasets of various sessions are denoted by $\{\boldsymbol{D}_0^{tr}, \boldsymbol{D}_1^{tr}, \ldots, \boldsymbol{D}_m^{tr}, \ldots, \boldsymbol{D}_{M-1}^{tr}\}$ and $\{\boldsymbol{D}_0^{te}, \boldsymbol{D}_1^{te}, \ldots, \boldsymbol{D}_m^{te}, \ldots, \boldsymbol{D}_{M-1}^{te}\}$, respectively. $\boldsymbol{D}_m^{tr}$ and $\boldsymbol{D}_m^{te}$ have the same label set which is denoted by $\boldsymbol{L}_m$. The dataset in different sessions do not have the same type of classes, namely $\forall m, n$ and $m \neq n$, $\boldsymbol{L}_m \cap \boldsymbol{L}_n = \emptyset$. In the $m$th session, only $\boldsymbol{D}_m^{tr}$ can be used to train the classification system, and the trained model is assessed on the testing dataset of current and all prior sessions, namely $\boldsymbol{D}_0^{te} \cup \boldsymbol{D}_1^{te} \ldots \cup \boldsymbol{D}_m^{te}$. The dataset $\boldsymbol{D}_0^{tr}$ is a large-scale dataset which has adequate samples per class for model training. In contrast, each one of the datasets $\boldsymbol{D}_1^{tr}$ to $\boldsymbol{D}_{M-1}^{tr}$ is a small-scale dataset, and is composed of $N$ classes ($K$ training samples per class). Namely, the training dataset $\boldsymbol{D}_m^{tr}$ in the $m$th ($1 \leq m \leq (M-1)$) incremental session is divided into many $N$-way $K$-shot training subsets.

## 2.2. The framework

As shown in Figure 1, the framework of our method includes base and incremental sessions. There are three steps in the base session, including training EE, training SC, and producing prototypes of base classes. In incremental session, the SC is continually expanded (trained) to recognize new classes.

In the base session, we first train an EE in a supervised way on $\boldsymbol{D}_0^t$ which has abundant samples. Then, the EE is frozen for alleviating catastrophic forgetting and is used to extract embeddings from Log Mel-spectra of samples. Inspired by the success of ResNet [37] for extracting embedding in audio and visual processing tasks [38], [39], we use the backbone of a ResNet18 as the EE here. Next, $\boldsymbol{D}_0^t$ is split into many batches. Each batch consists of a support set and a query set. The SC is episodically trained on each batch using the loss

$$\mathcal{L}(\boldsymbol{x}, y; \alpha) = -log(e^{cos(f_\alpha(\boldsymbol{x}),\hat{\mu}_y)}/\sum_{h=0}^{|L|} e^{cos(f_\alpha(\boldsymbol{x}),\hat{\mu}_h)}) \quad (1)$$

where $\boldsymbol{x}$ and $y$ are input sample and its corresponding label; $\alpha$ denotes the parameters of EE; $\boldsymbol{L} = \boldsymbol{L}_0 \cup \boldsymbol{L}_1 \cdots \cup \boldsymbol{L}_m$; $\hat{\mu} = \mu + \mathcal{N}(0,1) \odot \sigma$ denotes the weight of SC; and $f_\alpha(\boldsymbol{x})$ denotes the embedding of $\boldsymbol{x}$. $\mu$ and $\sigma$ stand for the mean vector and variance vector of the SC. To keep the stability of SC, the mean vectors of embeddings belonging to the same kind of class are used to initialize $\mu$. $cos(\cdot,\cdot)$ denotes the cosine similarity.

In the $m$th incremental session, embeddings of support set in $\boldsymbol{D}_m^{tr}$ are first extracted. Next, embeddings of samples in current session and prototypes of all previous sessions are used to train the SC with a joint loss

$$\mathcal{L}_j^m = (1-\lambda)\mathcal{L} + \lambda \mathcal{L}_p \quad (2)$$

where $\lambda$ is an adjustable coefficient; $\mathcal{L}$ denotes the loss defined by Eq. (1), and $\mathcal{L}_p$ stands for prototype loss. $\mathcal{L}_p$ is defined by

$$\mathcal{L}_p(p_c, c) = -log\left(\frac{e^{cos(p_c,\hat{\mu}_c)}}{\sum_{h=0}^{|L|} e^{cos(p_h,\hat{\mu}_h)}}\right), \quad (3)$$

where $c \in (\boldsymbol{L}_0 \cup \boldsymbol{L}_1 \cdots \cup \boldsymbol{L}_{m-1})$ denotes the label of prototypes and $p_c$ is the prototype of class $c$. Finally, mean vectors of the trained SC are used as the updated prototypes for evaluation.

## 2.3. Stochastic classifier

In this study, cosine similarity between the embeddings and the prototypes is used to calculate the class score for a specific embedding. For a given input sample $\boldsymbol{x}$ from class $C_i$, its embedding is denoted by $f_\alpha(\boldsymbol{x})$ where $\alpha$ stands for the parameters of the EE $f_\alpha(\cdot)$. The prototype of a deterministic classifier corresponding to class $C_i$ is represented by $\mu_i^d$. The cosine similarity between the embedding $f_\alpha(\boldsymbol{x})$ and the prototype $\mu_i^d$ is computed by

$$\cos(f_\alpha(\boldsymbol{x}), \mu_i^d) = \frac{f_\alpha(\boldsymbol{x}) \cdot \mu_i^d}{\|f_\alpha(\boldsymbol{x})\|_2 \cdot \|\mu_i^d\|_2}, \quad (4)$$

where $\|\cdot\|_2$ denotes 2-norm. Figure 2 (a) shows the embedding $f_\alpha(\boldsymbol{x})$ and two prototypes, $\mu_j^d$ and $\mu_i^d$. The grey shaded area represents the region where $f_\alpha(\boldsymbol{x})$ can be correctly classified to class $C_i$, and $b_{ij}$ denotes the classification boundary between these two classifiers (represented by prototypes $\mu_i^d$ and $\mu_j^d$).

Using a SC $f_\beta = \{\mu, \sigma\}$ at the classification head is similar to the use of multiple classifiers, where $\mu$ and $\sigma$ stand for the mean vector and variance vector of the classifier. For a given sample $\boldsymbol{x}$, the class score of the SC is proportional to the cosine similarity $\cos(f_\alpha(\boldsymbol{x}), \hat{\mu})$ which is defined by

$$\cos(f_\alpha(\boldsymbol{x}), \hat{\mu}) = \frac{f_\alpha(\boldsymbol{x}) \cdot \hat{\mu}}{\|f_\alpha(\boldsymbol{x})\|_2 \cdot \|\hat{\mu}\|_2} \quad (5)$$

where $\hat{\mu} = \mu + \mathcal{N}(0,1) \odot \sigma$ is used to sample the prototypes. As shown in Figure 2 (b), classification boundary turns into a margin determined by $f_\beta^j$ and $f_\beta^i$. When $f_\alpha(x)$ locates in this margin, we can only have an inexplicit prediction that $f_\alpha(x)$ maybe belongs to class $j$, which means more powerful punishment is put into effect in next training iteration and enforce the EE to produce more informative embeddings in base session. Moreover, the SC allows distinguishing between the epistemic uncertainty and the aleatoric uncertainty [45]. Benefiting from the distinguishing feature, new class will have high epistemic uncertainty due to scarcity of training samples. Then, the problem of overfitting in new class will be alleviated.

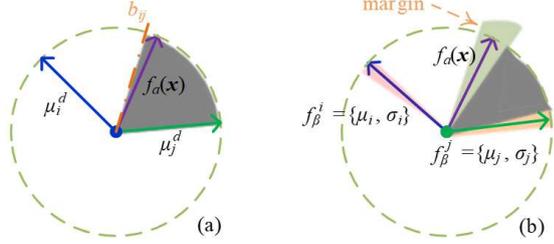

Figure 2: *The classification boundary between two classes in (a) deterministic classifier, (b) stochastic classifier. The margin in (b) leads to more discriminative classification boundaries.*

## 3. Experiments

### 3.1. Experimental datasets

Experiments are done on the datasets chosen from two audio corpora, including NSynth [41] and LibriSpeech [42]. These two audio corpora can be publicly accessed and are widely used in prior works. There are 306,043 audio snippets in the NSynth, and each audio snippet is of four seconds to represent one type of musical instruments. There are 1,006 musical instruments (classes). The LibriSpeech is a speech corpus of approximately 1,000 hours of audiobooks uttered by 2,484 speakers (classes).

The datasets chosen from the NSynth and LibriSpeech, are labeled as NS-100 and LS-100, respectively. These datasets are split into two parts without overlaps of classes, namely $D_0$ and $D_m$ ($1 \leq i \leq (M-1)$). $D_0$ (or $D_m$) is composed of training dataset $D_0^{tr}$ (or $D_m^{tr}$) and testing dataset $D_0^{te}$, (or $D_m^{te}$).

In each episodic training step, $N \cdot K$ samples are randomly chosen from $N$ classes ($K$ samples per class) in the training dataset to generate support set, and then $N \cdot K$ different samples of the same $N$ classes ($K$ samples per class) are randomly chosen from the training dataset to construct the query set. The selections of classes and samples per class are repeated until all classes and samples in the training dataset are chosen once. The samples in various batches are different to each other. In testing step, all testing datasets are input to the EE and classifier as a whole. Table 1 lists the detailed information of NS-100/LS-100.

Table 1: *Detailed information of NS-100/LS-100.*

| Parameters | $D_0$ | | $D_m$ | |
|---|---|---|---|---|
| | $D_0^{tr}$ | $D_0^{te}$ | $D_m^{tr}$ | $D_m^{te}$ |
| #Classes | 55/60 | 55/60 | 45/40 | 45/40 |
| #Samples | 11000/300000 | 5500/6000 | 4500/20000 | 4500/4000 |
| Length (hours) | 12.23/16.66 | 1.52/3.33 | 5.00/11.11 | 5.00/2.22 |
| #Samples/Class | 200/500 | 100/100 | 100/500 | 100/100 |

#Samples/class: number of samples per class. Numbers of the left and right of the slash are corresponding values of NS-100 and LS-100, respectively.

### 3.2. Experimental setup

Accuracy is defined as the number of correctly classified samples divided by total number of samples involved in classification. It is used to measure the performance of various methods in each session. Both AA and PD are used to assess overall performance of different methods, and defined by

$$AA = \frac{1}{M}\sum_{m=0}^{M-1} A_m, \quad (6)$$

$$\begin{cases} PD = A_0 - A_{M-1}, & \text{for the classes of Base and All} \\ PD = A_1 - A_{M-1}, & \text{for the classes of Incremental} \end{cases} \quad (7)$$

where $A_m$ stands for the accuracy in session $m$. Frame length and frame overlapping are 25 ms and 10 ms, respectively. Dimensions of log Mel-spectrum and prototypes are 128 and 512, respectively. The coefficient $\lambda$ is experimentally set to 0.9 and 0.6 for LS-100 and NS-100, respectively.

### 3.3. Experimental results

In the first experiment, we assess the effectiveness of our method with two classifiers on LS-100. The results obtained by our method with stochastic or deterministic classifiers are listed in Table 2. When SC is used, our method obtains higher AA scores for the classes of Base, Incremental and All (Base + Incremental). That is, SC outperforms deterministic classifier when they are adopted in our method for audio classification on LS-100.

Table 2: *Average accuracy scores obtained by our method with deterministic classifier and stochastic classifier on LS-100.*

| Classes | Deterministic classifier | Stochastic classifier |
|---|---|---|
| Base | 91.88% | 92.13% |
| Incremental | 74.52% | 77.41% |
| All | 87.37% | 88.39% |

In the second experiment, we discuss the influence of the values of $N$ and $K$ on the performance of our method. The accuracy of the last session obtained by our method on LS-100 is given in Figure 3, from which three observations can be obtained. First, when $(N, K)$ is equal to (20, 20), our method obtains the highest accuracy score of 88.48%. Second, for the same number of ways, the larger the number of shots, the higher the accuracy scores. It is probably that with the increase of shots, the model acquires more information about new classes and obtains higher accuracy scores. Third, for the same number of shots, when the number of ways is equal to 20, our method obtains the highest accuracy score (except 5-way 1-shot). When the value of $N$ deviates from 20, accuracy scores decrease. The possible reasons are as follows. When the number of ways decreases, the number of incremental sessions will increase and the old classes are more likely to be forgotten. When the number of ways increases, the number of new classes in one incremental session will increase and the confusions between new classes in each session is more likely to increase.

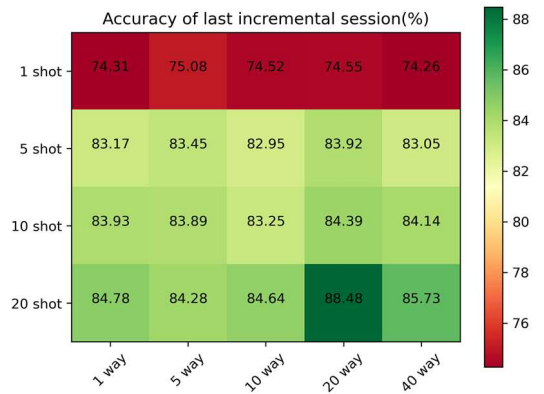

Figure 3: *Accuracy of the last session obtained by our method on LS-100.*

Table 3: *Results obtained by different methods on LS-100. Incr. denotes incremental classes.*

| Methods | Session | 0 (0-59) | 1 (60-64) | 2 (65-69) | 3 (70-74) | 4 (75-79) | 5 (80-84) | 6 (85-89) | 7 (90-94) | 8 (95-99) | AA (%) | PD (%) |
|---|---|---|---|---|---|---|---|---|---|---|---|---|
| Finetune | Base | 92.02 | 72.90 | 37.03 | 28.12 | 20.75 | 14.45 | 5.70 | 3.23 | 0.27 | 30.50 | 91.75 |
| | Incr. | - | 86.60 | 31.50 | 28.87 | 25.45 | 24.24 | 18.17 | 13.46 | 11.80 | 30.01 | 74.80 |
| | All | 92.02 | 73.95 | 36.24 | 28.27 | 21.93 | 17.33 | 9.86 | 7.00 | 4.88 | 32.39 | 87.14 |
| iCaRL | Base | 92.02 | 80.80 | 73.18 | 58.45 | 26.95 | 16.93 | 32.58 | 29.53 | 26.38 | 48.54 | 65.64 |
| | Incr. | - | 58.00 | 67.10 | 57.40 | 20.05 | 16.48 | 30.33 | 26.83 | 28.95 | 38.14 | 29.05 |
| | All | 92.02 | 79.05 | 72.31 | 58.24 | 25.23 | 16.80 | 31.83 | 28.54 | 27.41 | 47.94 | 64.61 |
| DFSL | Base | 91.93 | 91.93 | 91.88 | 91.85 | 91.83 | 91.86 | 91.85 | 91.85 | 91.84 | 91.87 | 0.09 |
| | Incr. | - | 53.60 | 61.90 | 50.67 | 48.90 | 51.56 | 47.97 | 44.11 | 45.38 | 50.51 | 8.22 |
| | All | 91.93 | 88.97 | 87.60 | 83.61 | 81.11 | 80.01 | 77.22 | 74.26 | 73.25 | 81.99 | 18.68 |
| CEC | Base | 91.72 | 91.67 | 91.25 | 91.14 | 91.10 | 91.07 | 90.97 | 90.66 | 90.72 | 91.14 | 1.00 |
| | Incr. | - | 86.30 | 82.76 | 69.67 | 68.25 | 67.06 | 66.03 | 60.35 | 60.05 | 70.06 | 26.25 |
| | All | 91.72 | 91.25 | 90.04 | 86.84 | 85.38 | 84.01 | 82.65 | 79.49 | 78.45 | 85.54 | 13.27 |
| Ours | Base | 92.73 | 92.72 | 92.62 | 92.48 | 92.48 | 92.47 | 92.34 | 90.74 | 90.67 | 92.14 | 2.06 |
| | Incr. | - | 86.84 | 84.26 | 77.74 | 74.99 | 75.79 | 74.60 | 72.45 | 72.64 | 77.41 | 14.20 |
| | All | 92.73 | 92.27 | 91.42 | 89.53 | 88.10 | 87.56 | 86.43 | 84.00 | 83.45 | 88.39 | 9.28 |

Table 4: *Results obtained by different methods on NS-100.*

| Methods | Session | 0(0-54) | 1(55-59) | 2(60-64) | 3(65-69) | 4(70-74) | 5(75-79) | 6(80-84) | 7(85-89) | 8(90-94) | 9(95-99) | AA (%) | PD (%) |
|---|---|---|---|---|---|---|---|---|---|---|---|---|---|
| Finetune | Base | 99.96 | 88.91 | 85.41 | 80.36 | 72.51 | 45.24 | 59.31 | 48.53 | 50.68 | 53.28 | 68.42 | 46.68 |
| | Incr. | - | 38.75 | 30.25 | 36.96 | 37.54 | 28.95 | 27.24 | 22.30 | 20.58 | 19.00 | 29.06 | 19.75 |
| | All | 99.96 | 84.73 | 76.92 | 71.06 | 63.18 | 40.15 | 47.99 | 38.33 | 38.01 | 37.86 | 59.82 | 62.1 |
| iCaRL | Base | 99.98 | 98.42 | 99.25 | 98.40 | 94.56 | 82.36 | 85.09 | 80.59 | 75.78 | 74.53 | 88.90 | 25.45 |
| | Incr. | - | 36.94 | 31.88 | 35.03 | 38.33 | 35.27 | 30.76 | 26.75 | 25.52 | 22.27 | 31.42 | 14.67 |
| | All | 99.98 | 93.30 | 88.88 | 84.82 | 79.57 | 67.65 | 65.92 | 59.65 | 54.62 | 51.01 | 74.54 | 48.97 |
| DFSL | Base | 99.93 | 99.11 | 98.83 | 95.83 | 94.84 | 94.81 | 94.39 | 93.76 | 92.06 | 91.61 | 95.52 | 8.32 |
| | Incr. | - | 57.01 | 55.57 | 59.89 | 59.35 | 56.46 | 52.29 | 50.94 | 52.57 | 52.49 | 55.17 | 4.52 |
| | All | 99.93 | 96.00 | 92.95 | 89.26 | 86.47 | 83.66 | 80.28 | 77.68 | 76.12 | 75.01 | 85.74 | 24.92 |
| CEC | Base | 99.96 | 99.87 | 99.90 | 99.29 | 99.24 | 99.30 | 99.26 | 99.24 | 99.20 | 99.23 | 99.45 | 0.73 |
| | Incr. | - | 71.06 | 71.61 | 72.37 | 69.17 | 69.20 | 66.92 | 64.80 | 65.28 | 63.59 | 68.22 | 7.47 |
| | All | 99.96 | 97.47 | 95.56 | 93.52 | 91.22 | 89.90 | 87.85 | 85.84 | 84.92 | 83.19 | 90.94 | 16.77 |
| Ours | Base | 99.98 | 98.08 | 98.69 | 97.38 | 96.44 | 97.43 | 96.99 | 97.53 | 96.10 | 96.81 | 97.56 | 3.17 |
| | Incr. | - | 95.60 | 94.73 | 93.45 | 92.53 | 85.20 | 81.53 | 78.50 | 79.44 | 77.86 | 86.53 | 17.74 |
| | All | 99.98 | 97.88 | 98.08 | 96.53 | 95.55 | 93.61 | 91.54 | 90.13 | 89.09 | 88.29 | 94.07 | 11.69 |

In the third experiment, our method is compared with four baseline methods. These methods are marked as Finetune [43], iCaRL [44], DFSL [35], and CEC [32]. In the Finetune method, the model can quickly adapt to new classes after finetuning with training samples of new classes. Hence, the model is inclined to overfit the new classes and forget the old ones. In the iCaRL method, data retention and knowledge distillation are used to train EE and classifier. When new classes emerge continually, the model is inclined to gradually forget the old classes. In the DFSL method, an attention-based generator for expanding classifier's weights and a cosine-similarity based classifier are used for realizing FCAC. In the CEC method, continually evolved classifiers are adopted for identifying new classes and a graph model is designed to propagate the contextual information between classifiers for updating prototypes.

All baseline methods are realized using open-source codes, whose parameters are configured in accordance with the practices in the corresponding papers and optimally adjusted on the training data. In this experiment, ($N$, $K$) is set to (5, 5) without the loss of generality. Under the same conditions, the scores of accuracies, AA and PD obtained by various methods on LS-100 and NS-100 are given in Tables 3 and 4, respectively.

As shown in Tables 3 to 4, our method achieves AA scores of 88.39% and 94.07% for both base and incremental classes (the rows of All in the tables) on the LS-100 and NS-100, respectively. These AA scores are higher than the counterparts obtained by all baseline methods. Our method obtains PD scores of 9.28% and 11.69% for both base and incremental classes on the LS-100 and NS-100, respectively. These PD scores are lower than the counterparts achieved by all baseline methods. That is, our method exceeds all baseline methods in both AA and PD. The advantage of our method over all baseline ones in AA and PD is mainly because of the usage of SC for updating prototypes. The SC can effectively reduce confusions among the updated prototypes of different classes. Compared with the baseline methods, our method recognizes new classes better and forget old ones less.

## 4. Conclusions

In this paper, we discussed the FCAC problem, and tried to address it using a SC. Based on the description of our method and experimental evaluations, we can draw two conclusions. First, our method outperforms state-of-the-art methods in both AA and PD under the same conditions. Second, we designed a SC for updating prototypes in incremental sessions. The prototype vectors are sampled from a distribution instead of deterministic vectors.

Although our method possesses advantages over baseline ones, it still needs to be improved. For instance, the EE is not updated in each incremental session and thus its generalization ability needs to be boosted. The future work will consider updating the EE or jointly updating the EE and SC in incremental sessions.

## 5. Acknowledgements

This work was supported by the national natural science foundation of China (62111530145, 61771200), international scientific research collaboration project of Guangdong (2021A0505030003), and Guangdong basic and applied basic research foundation (2021A1515011454, 2022A1515011687).

# 6. References

[1] Y. Li, Y. Zhang, X. Li, M. Liu, W. Wang, and J. Yang, "Acoustic event diarization in TV/movie audios using deep embedding and integer linear programming," *Multimedia Tools Applications*, vol. 78, pp. 33999-34025, 2019.

[2] K. Drossos, S. Lipping, and T. Virtanen, "Clotho: an audio captioning dataset," in *Proc. IEEE ICASSP*, 2020, pp. 736-740.

[3] Y. Li, X. Li, Y. Zhang, M. Liu, and W. Wang, "Anomalous sound detection using deep audio representation and a BLSTM network for audio surveillance of roads," *IEEE Access*, vol. 6, pp. 58043-58055, 2018.

[4] A. Terenzi, N. Ortolani, I. Nolasco, E. Benetos, and S. Cecchi, "Comparison of feature extraction methods for sound-based classification of honey bee activity," *IEEE/ACM TASLP*, vol. 30, pp. 112-122, 2022.

[5] D. Barchiesi, D. Giannoulis, D. Stowell, and M.D. Plumbley, "Acoustic scene classification: classifying environments from the sounds they produce," *IEEE Signal Process. Mag.*, vol. 32, no. 3, pp.16-34, 2015.

[6] Y. Zeng, Y. Li, Z. Zhou, R. Wang, and D. Lu, "Domestic activities classification from audio recordings using multi-scale dilated depthwise separable convolutional network," in *Proc. of IEEE MMSP*, 2021, pp. 1-5.

[7] T. Iqbal, Y. Cao, Q. Kong, M. D. Plumbley, and W. Wang, "Learning with out-of-distribution data for audio classification," in *Proc. of IEEE ICASSP*, 2020, pp. 636-640.

[8] A. Politis, A. Mesaros, S. Adavanne, T. Heittola, and T. Virtanen, "Overview and evaluation of sound event localization and detection in DCASE 2019," *IEEE/ACM TALSP*, vol. 29, pp. 684-698, 2021.

[9] K. Choi, M. Kersner, J. Morton, and B. Chang, "Temporal knowledge distillation for on-device audio classification," in *Proc. of IEEE ICASSP*, 2022, pp. 486-490.

[10] M. Mohaimenuzzaman, C. Bergmeir, and B. Meyer, "Pruning vs XNOR-Net: A comprehensive study of deep learning for audio classification on edge-devices," *IEEE Access*, vol.10, pp.6696-6707, 2022.

[11] Y. Li, M. Liu, K. Drossos, and T. Virtanen, "Sound event detection via dilated convolutional recurrent neural networks," in *Proc. of IEEE ICASSP*, 2020, pp. 286-290.

[12] H. Shin, J. Kwon Lee, J. Kim, and J. Kim, "Continual learning with deep generative replay," in *Proc. of NIPS*, 2017, pp. 2990-2999.

[13] S. Zhang, Y. Qin, K. Sun, and Y. Lin, "Few-shot audio classification with attentional graph neural networks," in *Proc. of INTERSPEECH*, 2019, pp. 3649-3653.

[14] S. Chou, K. Cheng, J. R. Jang, and Y. Yang, "Learning to match transient sound events using attentional similarity for few-shot sound recognition," in *Proc. of IEEE ICASSP*, 2019, pp. 26-30.

[15] J. Pons, J. Serrà, and X. Serra, "Training neural audio classifiers with few data," in *Proc. of IEEE ICASSP*, 2019, pp. 16-20.

[16] Y. Wang, and D. V. Anderson, "Hybrid attention-based prototypical networks for few-shot sound classification," in *Proc. of IEEE ICASSP*, 2022, pp. 651-655.

[17] Y. Wang, J. Salamon, N. J. Bryan, and J. P. Bello, "Few-shot sound event detection," in *Proc. of IEEE ICASSP*, 2020, pp. 81-85.

[18] D. Yang, H. Wang, Y. Zou, Z. Ye, and W. Wang, "A mutual learning framework for few-shot sound event detection," in *Proc. of IEEE ICASSP*, 2022, pp. 811-815.

[19] J. Snell, K. Swersky, and R. Zemel, "Prototypical networks for few-shot learning," in *Proc. of NIPS*, 2017, pp. 4078-4088.

[20] C. Finn, P. Abbeel, and S. Levine, "Model-agnostic meta-learning for fast adaptation of deep networks," in *Proc. of ICML*, 2017, pp. 1-13.

[21] H.J. Chen, A.C. Cheng, D.C. Juan, W. Wei, and M. Sun, "Mitigating forgetting in online continual learning via instance-aware parameterization," in *Proc. of NeurIPS*, 2020, vol. 33, pp. 17466-17477.

[22] Y. Liu, B. Schiele, and Q. Sun, "Adaptive aggregation networks for class-incremental learning," in *Proc. of IEEE/CVF CVPR*, 2021, pp. 2544-2553.

[23] E. Koh, F. Saki, Y. Guo, C. Hung, and E. Visser, "Incremental learning algorithm for sound event detection," in *Proc. of IEEE ICME*, 2020, pp.1-6.

[24] X. Wang, C. Subakan, E. Tzinis, P. Smaragdis, and L. Charlin, "Continual learning of new sound classes using generative replay," in *Proc. of IEEE WASPAA*, 2019, pp. 308-312.

[25] D. Ma, C. I. Tang, and C. Mascolo, "Improving feature generalizability with multitask learning in class incremental learning," in *Proc. of IEEE ICASSP*, 2022, pp. 4173-4177.

[26] S. Karam, S.-J. Ruan, and Q.M.u. Haq, "Task incremental learning with static memory for audio classification without catastrophic interference," *IEEE Consumer Electronics Magazine*, vol. 11, no. 5, pp. 101-108, 2022.

[27] M.A. Hussain, C.-L. Lee, and T.-H. Tsai, "An efficient incremental learning algorithm for sound classification," *IEEE MultiMedia*, 2022, pp. 1-8. doi: 10.1109/MMUL.2022.3208923.

[28] X. Tao, X. Hong, X. Chang, S. Dong, X. Wei, and Y. Gong, "Few-shot class-incremental learning," in *Proc. of IEEE/CVF CVPR*, 2020, pp. 12180-12189.

[29] A. Ayub, and A.R. Wagner, "Cognitively-inspired model for incremental learning using a few examples," in *Proc. of IEEE/CVF CVPRW*, 2020, pp. 897-906.

[30] S. Gidaris, and N. Komodakis, "Dynamic few-shot visual learning without forgetting," in *Proc. of IEEE/CVF CVPR*, 2018, pp. 4367-4375.

[31] P. Mazumder, P. Singh, and P. Rai, "Few-shot lifelong learning," in *Proc. of AAAI*, 2021, pp. 1-9.

[32] C. Zhang, N. Song, G. Lin, Y. Zheng, P. Pan, and Y. Xu, "Few-shot incremental learning with continually evolved classifiers," in *Proc. of IEEE/CVF CVPR*, 2021.

[33] A. Cheraghian, S. Rahman, P. Fang, S. K. Roy, L. Petersson, and M. Harandi, "Semantic-aware knowledge distillation for few-shot class-incremental learning," in *Proc. of IEEE/CVF CVPR*, 2021.

[34] B. Yang, M. Lin, Y. Zhang, B. Liu, X. Liang, R. Ji, and Q. Ye, "Dynamic support network for few-shot class incremental learning," *IEEE TPAMI*, vol. 45, no. 3, pp. 2945-2951, 2023.

[35] Y. Wang, N. J. Bryan, M. Cartwright, J. Pablo Bello, and J. Salamon, "Few-shot continual learning for audio classification," in *Proc. of IEEE ICASSP*, 2021, pp. 321-325.

[36] Z. Lu, Y. Yang, X. Zhu, C. Liu, Y. -Z. Song and T. Xiang, "Stochastic classifiers for unsupervised domain adaptation," in *Proc. of CVPR*, 2020, pp. 9108-9117.

[37] K. He, X. Zhang, S. Ren, and J. Sun, "Deep residual learning for image recognition," in *Proc. of IEEE/CVF CVPR*, 2016, pp. 770-778.

[38] X. Shi, E. Cooper, and J. Yamagishi, "Use of speaker recognition approaches for learning and evaluating embedding representations of musical instrument sounds," *IEEE/ACM TASLP*, vol. 30, pp. 367-377, 2022.

[39] R. Qian, T. Meng, B. Gong, M.H. Yang, H. Wang, S. Belongie, and Y. Cui, "Spatiotemporal contrastive video representation learning," in *Proc. of IEEE/CVF CVPR*, 2021, pp. 6960-6970.

[40] D. Kingma, and J. Ba, "Adam: A method for stochastic optimization," in *Proc. of ICLR*, 2015, pp. 1-15.

[41] J. Engel, C. Resnick, A. Roberts, S. Dieleman, M. Norouzi, D. Eck, and K. Simonyan, "Neural audio synthesis of musical notes with WaveNet autoencoders," in *Proc. of ICML*, 2017, vol. 70, pp. 1068-1077.

[42] V. Panayotov, G. Chen, D. Povey, and S. Khudanpur, "Librispeech: An ASR corpus based on public domain audio books," in *Proc. of IEEE ICASSP*, 2015, pp. 5206-5210.

[43] P. Dhar, R. V. Singh, K.-C. Peng, Z. Wu, and R. Chellappa, "Learning without memorizing," in *Proc. IEEE/CVF CVPR*, 2019, pp. 5138-5146.

[44] S.-A. Rebuffi, A. Kolesnikov, G. Sperl, and C.H. Lampert, "iCaRL: Incremental classifier and representation learning," in *Proc. of IEEE CVPR*, 2017, pp. 5533-5542.

[45] L.V. Jospin, H. Laga, F. Boussaid, et al., "Hands-on Bayesian neural networks—A tutorial for deep learning users," *IEEE Computational Intelligence Magazine*, vol. 17, no. 2, pp. 29-48, 2022.